# Trajectory dependence of electronic energy-loss straggling at keV ion energies

S. Lohmann[1,2], R. Holeňák[1], P. L. Grande[3], D. Primetzhofer[1]

[1] *Department of Physics and Astronomy, Uppsala University, Box 516, 751 20 Uppsala, Sweden*

[2] *Institute of Ion Beam Physics and Materials Research, Helmholtz-Zentrum Dresden-Rossendorf e.V. (HZDR), Bautzner Landstr. 400, 01328 Dresden, Germany*

[3] *Physics Department, Institute of Physics, Federal University of Rio Grande do Sul (UFRGS), CEP 91501-970, Porto Alegre, RS, Brazil*

**Abstract**

We have measured the electronic energy-loss straggling of protons, helium, boron and silicon ions in silicon using a transmission time-of-flight approach. Ions with velocities between 0.25 and 1.6 times the Bohr velocity were transmitted through single-crystalline Si(100) nanomembranes in either channelling or random geometry to study the impact parameter dependence of energy-loss straggling. Nuclear and path length contributions to the straggling were determined with the help of Monte Carlo simulations. Our results exhibit an increase in straggling with increasing ion velocity for channelled trajectories for all projectiles as well as for protons and helium in random geometry. In contrast for heavier ions, electronic straggling at low velocities does not decrease further but plateaus and even seems to increase again. We compare our experimental results with transport cross section calculations. The satisfying agreement for helium shows that electronic stopping for light ions is dominated by electron-hole pair excitations, and that the previously observed trajectory dependence can indeed be attributed to a higher mean charge state for random trajectories. No agreement is found for boron and silicon indicating the breakdown of models based solely on electron-hole pair excitations, and that local electron-promotion and charge-exchange events significantly contribute to energy loss at low velocities.

**Introduction**

When ions travel through matter, they transfer energy to the sample atomic nuclei and electrons. This process is used in research and technology to modify and analyse materials [1] as well as in medicine for cancer therapy [2]. Whether you want to change a semiconductor's characteristics by ion implantation or destroy a tumour cell with minimal damage to the healthy tissue around it, precise, quantitative knowledge of the energy deposition is of utmost importance [3].

Quantification of the ion energy loss per path length is commonly done in the form of the stopping power $S=S_n+S_e$, where $S_n$ and $S_e$ correspond to the energy transfer to the atomic nuclei and the target electrons, respectively. Whereas the stopping power denotes a mean value, the energy transfer actually happens in many individual collisions between the ion and the sample constituents, i.e., it can be treated as a stochastic process. The ion beam at a certain sample depth is, therefore, no longer monoenergetic but exhibits a broader energy distribution, which can be characterised by its variance called energy-loss straggling [4]. Energy-loss straggling fundamentally limits the achievable energy resolution for ion beam analysis and the depth precision in ion implantation and irradiations. Like stopping it thus needs to be modelled accurately for optimized applications. Straggling of fast ions (i.e. with ion velocity $v \gg v_0$, where $v_0$ corresponding to 1 atomic unit (a.u.) denotes the Bohr velocity) was already theoretically described by Bohr [5], and his model was later extended to lower velocities [6–8]. For $v \sim v_0$, the ion velocity is comparable to the velocity of valence electrons in the solid, and interactions are no longer adiabatic. The dynamics and complexity of interactions in this regime make them an active field of fundamental experimental [9–12] and theoretical research [13–16]. Previously, some of us have studied the impact parameter dependence of electronic stopping of light and heavy ions with velocities $v \sim v_0$ and $< v_0$ in Si [17,18]. For protons our observation matches experiments at MeV energies, namely an increasing trajectory dependence with higher ion velocities attributed to increasing contributions of localised core-electron excitations. In contrast, we observe a reverse trend for all heavier ions: an increasing difference in the energy lost along random and channelled trajectories with decreasing ion velocity. We attribute the additional energy loss along random trajectories to charge-exchange and electron-promotion events only accessible in close collisions. These events can contribute in two ways to the ion energy loss. First, they alter the mean charge state of the ion along its trajectory. While ions in this velocity regime can capture an electron via Auger processes at any impact parameter possible while traversing solid matter [19], reionisation is only possible at sufficiently small impact parameters. Therefore, ions travelling along random trajectories can be expected to have a higher mean charge state, and consequently higher energy loss, than channelled ions. Second, the electron promotion occurring during close collisions is connected with a direct, local energy transfer of the ion to the solid [20,21]. By only observing the ion energy loss, no conclusions can be drawn which one of these processes dominates. However, the first process is connected to a large number of collisions with electrons along all the trajectory whereas the second one essentially consists of few events with relatively large energy transfer, which are additionally restricted to low interaction distances between ion and target nuclei. As a consequence, the statistical distributions of the resulting electronic energy loss can, for equivalent total electronic loss, be expected to differ. Therefore, a signature of the dominating energy-loss process might be found in the electronic straggling.

To answer this question, we analyse the electronic energy-loss straggling of H, He, B and Si ions transmitted through Si(100) nanomembranes. We measure the ion energy loss via the flight time, and account for nuclear and geometrical straggling with the help of Monte Carlo methods. In addition, we compare our experimental results with the Chu model and transport cross section calculations.

**Experimental technique**

The experimental studies of straggling were performed using self-supporting, single-crystalline Si(100) nanomembranes (Norcada "UberFlat") with nominal thicknesses between 50 nm and 200 nm. Sample areal densities were determined with Rutherford backscattering spectrometry (RBS) with an uncertainty of better than 6 % (7.5 % for the thinnest samples with nominal thickness of 50 nm). Additionally, time-of-flight elastic recoil detection analysis was performed to check for potential light impurities. The purity in the bulk was found to be very high. Light H, C and O impurities could be detected on both surfaces indicating the expected presence of a surface oxide and carbohydrates. Further details of the characterisation process can be found in [18].

Straggling was measured using the Uppsala time-of-flight medium-energy ion scattering (ToF-MEIS) set-up [22,23]. Beams of singly-charged H, He, B and Si were produced by a 350 kV Danfysik implanter. The beam is electrostatically chopped resulting in beam pulses with typical lengths between 1 and 3 ns and an angular divergence significantly better than 0.056°. The beam cross section is defined by sets of horizontal and vertical slits in the beam line and kept smaller than (1 x 1) mm$^2$. The pulsing together with the spatial restriction leads to a current arriving on the sample of 2-3 fA. The vacuum in the scattering chamber is found to be below 1 x 10$^{-8}$ mbar. No on-site sample preparation was performed prior to experiments.

Ions are transmitted through the foil of interest. Specifically, we use foils with thicknesses of 53 nm, 135 nm and 200 nm as obtained from the RBS areal densities and assuming bulk density. The sample holder is mounted to a six-axis goniometer allowing for precise positioning and aligning of the beam direction with regard to specific crystal axes and planes. 290 mm behind the sample a position-sensitive microchannel plate (MCP) (DLD120 from RoentDek [24]) is located to detect transmitted particles. The circular detector with diameter 120 mm covers scattering angles ±11.5° and a solid angle of 0.13 sr. Behind the MCP two perpendicular delay-lines are mounted to determine the position of the detected particle. The energy of a particle is determined via its flight time and is available for every pixel on the detector.

Figure 1 illustrates this experimental approach showing the spatial distribution of transmitted ions on the detector in the inset. In this example 100 keV Si$^+$ ions (corresponds to 0.38 a.u.) were transmitted through a 53 nm thick Si(100) nanomembrane, which was rotated by $\theta_x$=1° ($\theta_y$=2°) around the goniometer x-axis (y-axis) out of the <100> channel. The goniometer axes were initially aligned with the major crystal axes. The main plot shows energy-loss distributions for selected regions-of-interest (ROIs) with 1 mm radius as indicated by the black circles in the inset. The initial beam position is located at x = 0.5 mm and y = -3.5 mm, i.e., the full, orange curve in the main plot gives the energy-loss distribution for relatively straight trajectories in this case. The majority of ions, however, is steered into the nearby (110)-plane, which is visible as the bright yellow spot in the inset. Furthermore, the intensity distribution exhibits a ring-like structure of higher intensity on the detector. This intensity maximum for specific scattering angles can be interpreted as a form of rainbow scattering [25]. The energy-loss distributions exhibit a component of lower energy loss for planar channelled ions and a part of higher loss for more random trajectories as previously reported in [17] and [18] and also briefly discussed in the introduction.

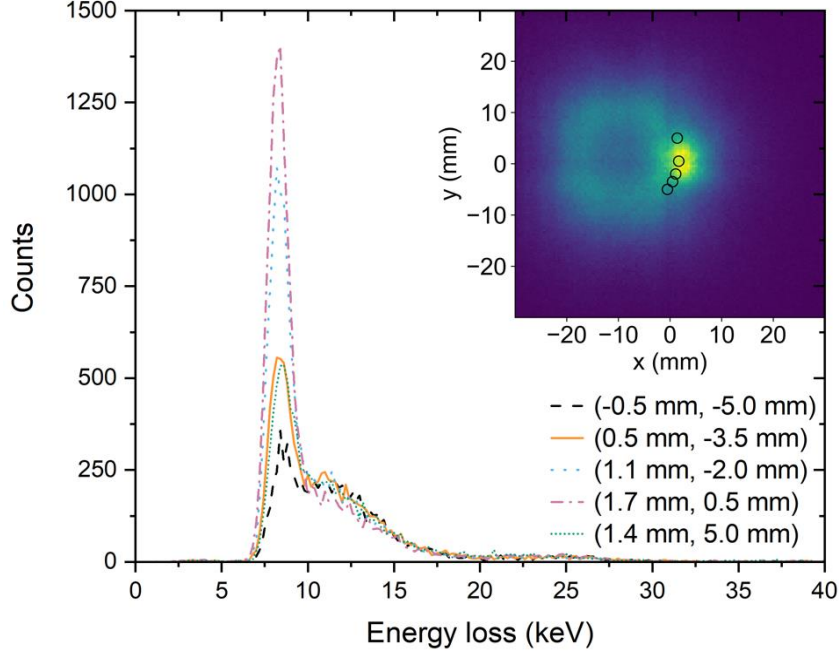

Figure 1: 100 keV Si$^+$ transmitted through a 53 nm thick self-supporting Si nanomembrane. The sample was turned out of channelling geometry by 1° (2°) around the x-axis (y-axis). The main plot shows energy-loss distributions that differ significantly even though the corresponding ion trajectories end in neighbouring ROIs on the detector. The figure legend states the ROI centres, and the ROI radius is 1 mm in all cases. The inset shows the spatial distribution of transmitted Si on the detector resulting in a ring-like structure of high intensity. The ROIs corresponding to the curves in the main plot are indicated by black circles. Note that only a part of the detector is shown.

**Data evaluation**

As can be already seen from Fig. 1, the observed shape of the energy distribution strongly depends on the chosen trajectory, and defining the distribution's variance might not be straight forward, in particular in the near proximity of a major channel. In the following we thus evaluate straggling for relatively straight trajectories both for channelling geometry and for samples rotated by at least 5° (10°) around the goniometer x-axis (y-axis). In this geometry the beam is no longer aligned with a low-index channel or plane, and we call it "random". We select trajectories ending in a ROI with radius 1 mm (corresponding to deflection angles ±0.2°). Figure 2 shows such energy-loss curves for the example of 40 keV B$^+$ (corresponds to 0.38 a.u.) transmitted through 53 nm Si(100). The blue open triangles give the measured energy loss in channel whereas the partly filled grey triangles belong to a random trajectory with $\theta_x$=5° and $\theta_y$=10°. The energy broadening of the as-measured curves $\delta E$ is not only caused by straggling but also by the finite detector resolution. For an ideal, initially monoenergetic beam $\delta E$ is described by

$$(\delta E)^2 = (\delta E_D)^2 + (\delta E_S)^2, \qquad (1)$$

where $\delta E_D$ and $\delta E_S$ denote the contributions of the detector resolution and straggling, respectively. $\delta E_S$ is the sum of nuclear ($\delta E_{S,n}$), electronic ($\delta E_{S,e}$) and geometrical straggling ($\delta E_{S,g}$). Note that since straggling depends on the number of interactions, it also depends on sample thickness. For simplicity we omitted the division by sample thickness in Eq. (1), but we present straggling values in the unit keV$^2$/Å. We are specifically interested in electronic straggling, therefore, all other contributions need to be subtracted from the measured distributions.

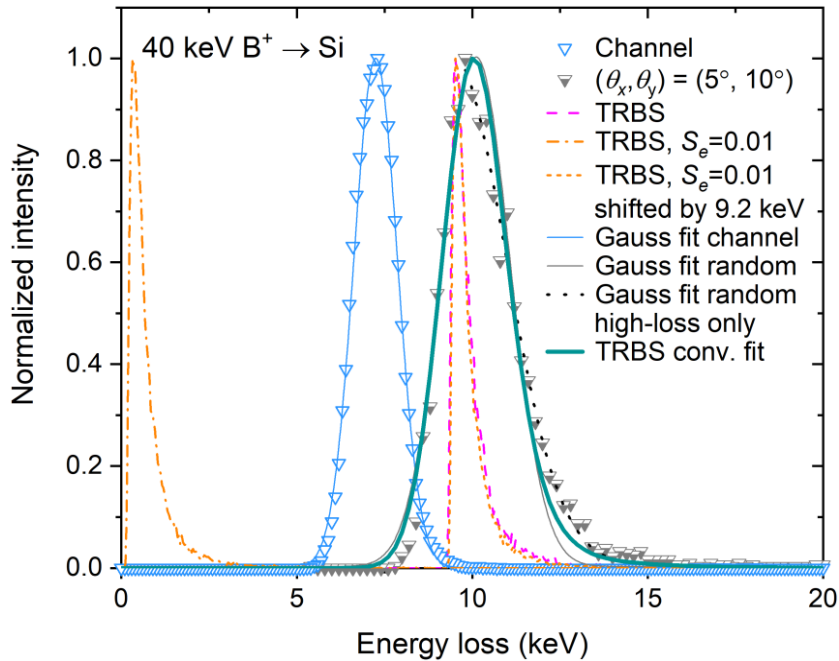

Figure 2: Energy-loss spectra for 40 keV B+ transmitted through 53 nm Si(100) both in channelling and random geometry. Also plotted are TRBS simulations both with full electronic stopping (fromTRIM85 [26]) and with electronic stopping set to 1 %. For better comparison, an energy-shifted version of the latter curve is shown as well. To demonstrate the straggling evaluation procedure, Gaussian functions are fitted to the experimental data. For the random spectrum, both a fit to the full peak and to the high-side are shown. In all cases the maximum intensity is normalized to 1.

The detector resolution of a ToF system is determined by the time resolution, which can be determined in two different ways in the Uppsala ToF-MEIS system. First, by measuring the flight time distribution of the direct beam on the detector by moving the sample out of the beam path. Note that the small difference between sample and transmission detector position (290 mm) is negligible compared to the flight distance behind the chopper (~8 m). This method provides a very direct assessment of the time resolution not hampered by any interactions of the beam. However, a potential disadvantage of this method is that the time resolution is not measured at the same time as the energy straggling in the sample. Therefore, the result is sensitive to a possible drift of the beam alignment with time. The second method makes use of ion-induced photon emission [27]. Photons are detected with the same MCP detector as transmitted ions and can be observed in every spectrum, thus, having the advantage of not requiring an additional measurement and not being susceptible to drift. For prompt photon emission, the width of the photon peak equals the time resolution of the system. For both methods we fit a Gaussian to the respective time distribution and take the standard deviation as the time uncertainty. By assuming a $E = 1/2\, m\, d^2/t^2$ relation between the energy $E$ of an ion with mass $m$ and the flight time $t$ for a fixed flight path $d$, we then calculate $\delta E_D$ via error propagation. The difference between the two methods is found to be much smaller than other uncertainties of the straggling evaluation. We, therefore, do not distinguish between the two methods.

To estimate the contributions of nuclear and geometrical straggling, we have performed simulations with the TRBS (TRIM for BackScattering) Monte Carlo package as a function of transmission angle [28]. TRBS is based on the popular TRIM code [29], but only explicitly calculates scattering events with a scattering angle larger than an adjustable cut-off angle thus allowing to simulate plural and multiple scattering events relevant at low energies [30]. Small-angle collisions are accounted for globally with a typical precision of 3 % [28]. The TRBS code has been shown to reproduce the multiple scattering background found in MEIS experiments with high accuracy [31]. We have chosen a cut-off angle of 0.1°, resulting in significantly more than 10 explicitly calculated scattering events per trajectory, but have found that the simulated results converge for angles <0.5°. We used the TRBS code with the Molière potential and Universal screening. TRBS does by default not model electronic straggling, therefore, the broadening of the simulated distribution can be entirely attributed to nuclear and geometrical straggling. The software allows to modify electronic stopping by a simple multiplicative constant. By setting this constant to 0.01, path length effects of electronic interactions can be suppressed. The resulting simulation only contains nuclear straggling. By performing two simulations for each ion type/energy, we could therefore determine the nuclear and the geometrical straggling component. For 40 keV $B^+$ the two TRBS simulations for a transmission angle of 0.8° and a Si thickness of 53.6 nm are plotted in Fig. 2. The dashed magenta curve depicts a simulation using the standard TRBS electronic stopping from TRIM85 [26] whereas the dash-dotted orange curve shows the simulation with this standard value multiplied by 0.01. For easier comparison we also show the latter curve shifted in energy (dotted orange curve). The used sample thickness includes the path length correction caused by the rotation of the sample with respect to the beam direction. We would also like to point out that TRBS simulations give the same energy-loss spectra for transmission angles smaller than 1°, which is much larger than the experimental angular acceptance of ±0.2°.

Channelled ions travel through the solid at large impact parameters and, thus, undergo virtually no elastic collisions. The restriction to a channel with small angular width also means that path length differences are negligible. We, therefore, assume that the straggling in channelling is purely electronic, and do not subtract any nuclear or geometrical contributions. Energy distributions obtained for channelled trajectories are symmetric and rather narrow for all ion species and velocities. We, therefore, fit the distributions by a Gaussian function with variance $\delta E$ as exemplarily shown by the blue full line in Fig. 2. Electronic straggling is then calculated according to Eq. (1).

The magnitude of nuclear stopping and straggling along random trajectories is dependent on the atomic number, and both are higher for heavier ions. For H and He, we assume that nuclear contributions are small, and thus determine electronic straggling as for channelled ions. For B, electronic and nuclear stopping from SRIM [32] are shown on the right axis of Fig. 3. The asterisks denote nuclear straggling along straight trajectories obtained from TRBS simulations performed as described above. As in [31], due to the high trajectory selectivity of our experimental approach, the nuclear stopping is found strongly reduced. Note that for the same reason geometrical straggling contributions are very small and significantly smaller than nuclear ones as can be seen by comparing the magenta and orange curves in Fig. 2. Nevertheless, we correct for nuclear and geometrical straggling for B as well as for Si ions.

Along random trajectories, the asymmetry of the measured energy-loss distribution depends both on ion velocity and type. For H, distributions also in random are almost perfectly symmetric. For He, a very small tail at high energy losses at the lowest velocities can be observed, however, the overall fit with a Gaussian function is satisfactory. As can be observed in Fig. 2, this is no longer true for heavier ions, and the shown random distribution is rather asymmetric with a tail towards high energy losses. The observed long tails in the TRBS simulations are known to arise from rare large-angle nuclear scattering events. Simply fitting a Gaussian to obtain the respective variance therefore leads to a rather bad fit as illustrated by the grey full line in Fig. 2. To demonstrate this asymmetry we have additionally fitted a Gaussian to the high-loss tail (the black dotted line in Fig. 2). Note that one could alternatively analyse skewness in addition to variance. We then calculated full peak and high-loss total straggling, i.e., still including nuclear effects, according to Eq. (1). To now subtract nuclear and geometrical straggling contributions, we convolute the full TRBS calculation (with full electronic stopping) with a Gaussian function. The result of the convolution is fitted to the experimental data with the variance and position of the Gaussian as fitting parameters. An example is shown as the thick petrol coloured line in Fig. 2. The variance minus $\delta E_D$ then gives the electronic straggling.

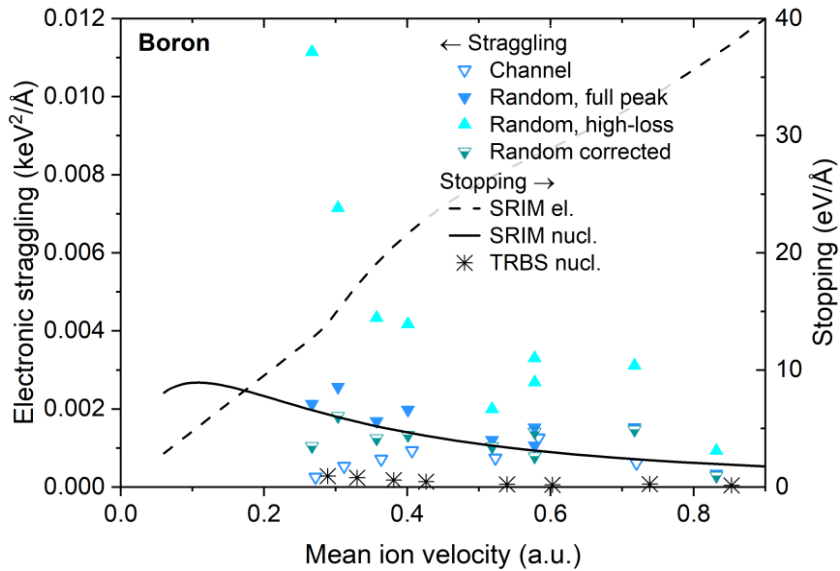

Figure 3: Left axis: Comparison of different methods to determine energy-loss straggling from experimental ToF-to-energy converted spectra using $B^+$ transmitted through Si(100) nanomembranes in channel and random. Data for random is analysed both by assuming a fully Gaussian distribution and by only fitting a Gaussian to the high-loss side of the distribution. The data set labelled "Random corrected" is corrected for nuclear and geometrical straggling contributions with the help of TRBS calculations. Right axis: Electronic and nuclear stopping of B in Si as obtained from SRIM. Asterisks give nuclear stopping along straight trajectories as obtained from TRBS simulations.

Figure 3 (left axis) shows results of the different evaluation methods as described above for $B^+$ ions transmitted through Si(100). Since straggling is velocity dependent we have decided to plot it as a function of the mean velocity in the target, which is calculated from the incident velocity and the exit velocity corresponding to the maximum of the measured energy distribution. Note that all given straggling values, therefore, are also mean values across a finite velocity bin. Data points labelled "Channel" (open triangles) and "Random" give the straggling as obtained from Eq. (1). "Full peak" and "High-loss" denote to which part of the experimental distribution the Gaussian was fitted to obtain $\delta E$. Corrections for nuclear and geometrical straggling are included in the half-filled petrol coloured data points.

**Results**

In the following we present experimental values of electronic energy straggling analysed as in detail described above for H, He, B and Si ions transmitted through thin, self-supporting Si(100) foils. We compare our experimental data with the Chu model [8] as an example for a commonly employed model, and show calculations based on the Transport Cross Section (TCS) and Penn-TCS approach. All these models (Chu, TCS and Penn-TCS) are based on a free electron gas (FEG) to describe the energy-loss straggling of ions in solids. The TCS model assumes a homogenous electron distribution described by the average valence electron density. Contrary to stopping calculations, straggling must be corrected by the Pauli principle [33], which is important at the present low ion energies. The key parameter for TCS calculations is the ion-electron potential, which is determined either from DFT calculations [34] or from a Yukawa potential with screening determined from the Friedel sum rule [35]. Although the description of the target is poor in this model, the ion-target interaction is nonlinear, going beyond the first order Born approximation. For real solids, the valence band electrons are better described by an inhomogeneous electron gas system or by a superposition of electron gases with different local densities. This local-density approximation concept was used by many authors, particularly in the Chu model, to evaluate the straggling using Lindhard's dielectric formalism. Selau et al. recently proposed a non-linear scheme based on an extension of the dielectric function model to describe the energy-loss straggling of ions in solids [36]. In this method, the energy-loss straggling is calculated for a statistical ensemble of FEGs with different plasmon energies as suggested by Penn in the context of the electron inelastic mean free path [37]. For each density, the momentum transfer rate from the electrons to the ion (the transport cross section) is calculated using a self-consistent screened electron-ion potential, which provides a nonlinear method to calculate stopping and straggling values. The present calculations of the Penn-TCS straggling were performed as in [36] with Electron Loss Function (ELF) from [38].

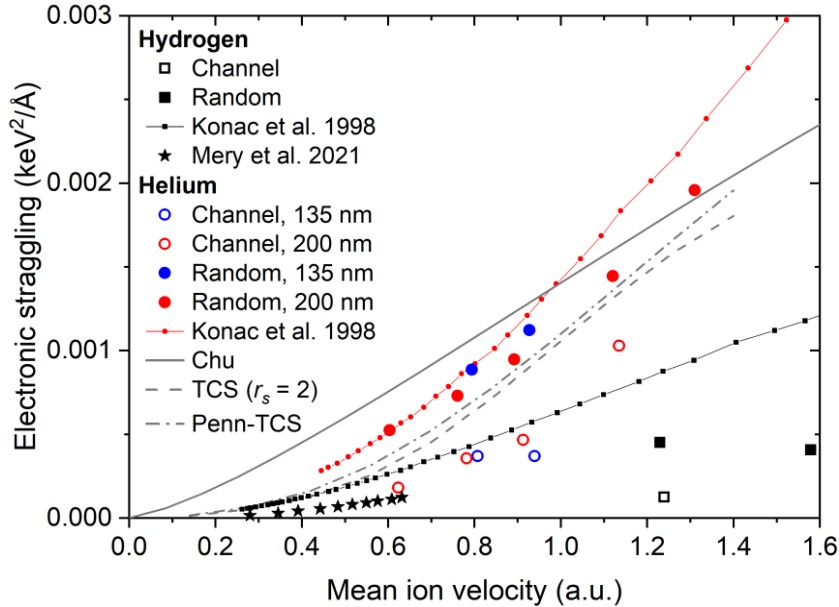

Figure 4: Electronic energy straggling of $H^+$ and $He^+$ transmitted through self-supporting Si(100) foils for channelling (open symbols) and random geometry (full symbols). For He, data for two different sample thicknesses is given (red and blue circles), H data has been obtained using the 200 nm thick foil (black squares). Also shown is experimental data from [39] for both ions types and from [40] for H only. For He we additionally present theoretical calculations as grey lines: Chu (full), TCS (dash) and Penn-TCS (dash-dot).

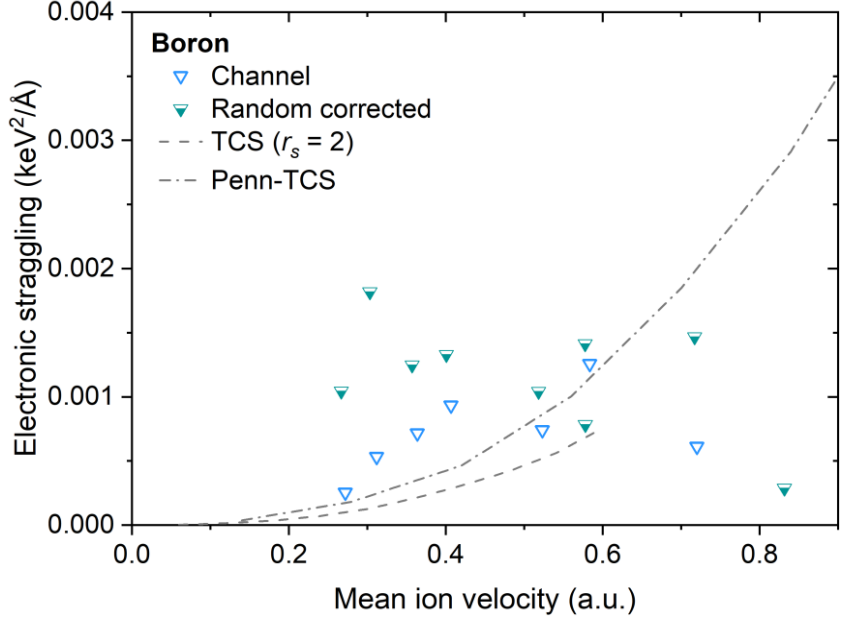

Figure 5: Electronic energy straggling of B$^+$ in Si(100) as already presented in Fig. 3. Here, we only show the channelled data set and electronic straggling along random trajectories corrected for nuclear and geometrical straggling. Experimental data is compared to straggling according to TCS (dash) and Penn-TCS (dash-dot) calculations.

Figure 4 shows the results for H$^+$ and He$^+$ projectiles. For He we present data from samples with two different thicknesses to check for effects of the path length (red and blue circles). The H data (black squares) was measured using the 200 nm thick foil only. For both ion types experimental data from [39] is plotted as symbols interconnected by lines. The authors of [39] performed a fit to data obtained from spectral line widths, which is also what we present here. The uncertainty is reported to be 4-8 %. Straggling data for H from [40] is shown as black stars. For H and He projectiles, straggling in channel and random has different values, but follows a similar velocity dependence, i.e., an increase of electronic straggling with velocity. The experimental data for He from [39] agrees well with our data at low velocities but lies higher at velocities greater than ~1 a.u. No actual energy distributions but only the fit to the data are shown in [39], and it is, therefore, not clear if their data is asymmetric. For H, our data lies significantly below the one from [39]. However, the data recently measured by Mery et al. at low ion velocities [40] also is lower and would approximately agree with our data, if extended to higher ion velocities. For He, theoretical models according to the Chu, TCS and Penn-TCS model are shown as grey lines.

To also compare the experimental data for B with theoretical models, we replot some of the data from Fig. 3 in Fig. 5. Since we are mainly interested in electronic straggling, only the random data corrected for nuclear and geometrical straggling is shown. As the Chu model was mainly developed for protons and He ions, we only present predictions from TCS and Penn-TCS models here. Velocity scaling of channelled data for B behaves in a similar way than for lighter ions with the exception of the data point at 0.72 a.u., which is, however, subject to a larger uncertainty due to the decreasing time resolution for faster ions (see discussion below). The random data, on the other hand, seems to be almost constant within the experimental uncertainty, and possibly slightly increasing towards lower velocities.

As an example for an even heavier projectile ion, we show the electronic straggling of Si$^+$ in Si(100) in Fig. 6. Again, we have measured straggling in channel and random orientation, and have applied corrections to the latter data as explained above. Penn-TCS calculations (dash-dot) were performed, and the results are best visible in the inset showing low straggling values only. Straggling in channel increases with increasing velocity as seen before for lighter ions. Along random trajectories, however, straggling seems to reach a minimum at around 0.3 a.u. to increase again towards lower velocities. The difference in magnitude between straggling in channelling and random orientation is also much greater than for lighter ions, and amounts to more than a factor 50. Note that the experimentally observed difference in electronic stopping only amounts to a factor 4 at the lowest velocity, and is even lower at higher velocities [18].

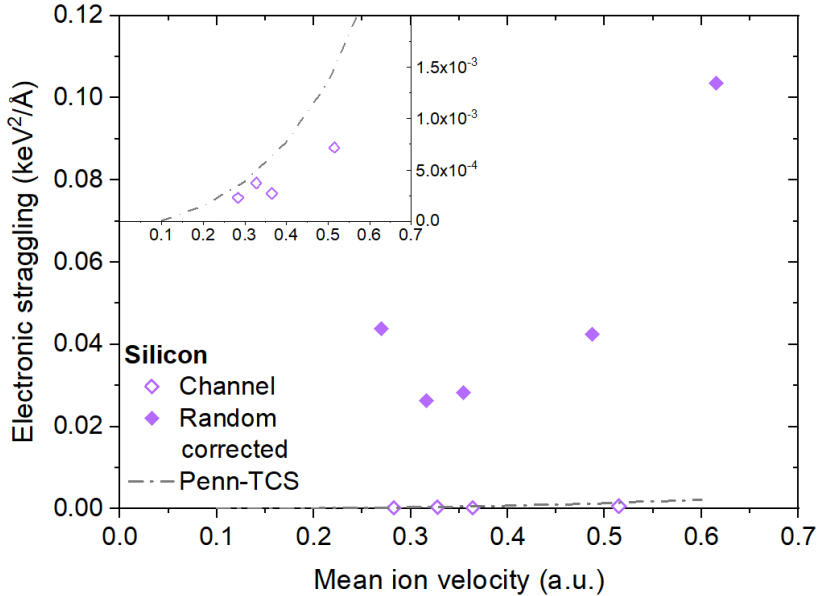

Figure 6: Electronic energy straggling of Si$^+$ in Si(100) for channelling (open diamonds) and random geometry (diamonds). For the latter, contributions from nuclear and geometrical straggling were subtracted as described in the text. Also shown are Penn-TCS calculations as a dash-dotted line. The inset is a magnification showing only small straggling values.

**Discussion**

*Influence of the experimental approach*

Electronic energy-loss straggling in Si exhibits a strong trajectory-dependence for all studied ions. Figure 1 demonstrates how ion trajectories in a single-crystal are influenced by the long-range order of the sample and experience channelling, blocking and rainbow scattering. For heavier ions, critical angles for channelling become larger compared to lighter ions of the same velocity. Channelling through high-index channels and planes is therefore more likely to happen and random trajectories are increasingly difficult to achieve in an experiment. The example shown in Fig. 1 is for Si ions, the heaviest projectiles studied in this work. The dashed black and the full orange curve are obtained from ROIs with centres only 1.8 mm apart. Still the resulting energy distributions look very different demonstrating how important it is to define the trajectory along which straggling is determined. As another example, for B we show two data points at the same velocity (0.58 a.u.) and identical random energy loss but slightly different alignment (cf. Fig. 3). The difference in observed straggling gives an estimate of the uncertainty caused by beam-sample alignment. In conclusion, energy-loss straggling is even more sensitive to the ion trajectory than energy loss.

It is an inherent characteristic of ToF-systems that for constant time resolution and flight distance, the energy resolution decreases for faster particles. For particles faster than a certain velocity, the straggling becomes small against the detector resolution so that the energy straggling can no longer be reliably determined via the ToF-method. Since straggling is much smaller for channelling than for random trajectories, the highest ion velocity, for which straggling can be measured, is lower in channel. Therefore, we e.g. show only one data point for straggling of H in channel.

We consider the sensitivity to the alignment between ion beam and crystal axis as the greatest contribution to our experimental uncertainty. The scatter of the already mentioned B data points at 0.58 a.u. gives a good estimate of this uncertainty. We therefore refrain from adding additional error bars. The good agreement between our data for He projectiles along random trajectories and previous experiments is a strong indication of the suitability of our experimental approach.

*Correlation between electronic and nuclear effects*

While electronic and nuclear straggling can potentially be entangled, it is not well known how much they influence each other in the velocity regime at hand. In particular, this correlation can be expected to differ for different ion-target combinations. TRIM and by extension TRBS is based on the binary collision approximation where nuclear and electronic scattering events are independent from each other. Thus, our approach, i.e., the convolution of the nuclear straggling contribution obtained from TRBS with a Gaussian representing the electronic straggling, gives a non-correlated result. Note that Fig. 3 reveals that nuclear straggling effects for B only start to play a role at velocities below 0.5 a.u. Apart from this exception the following observations are true for all data points corrected for nuclear straggling. In general, an asymmetry of the random energy-loss distribution matches the expectations from calculations, i.e., the existence of a few events with large energy loss correlated to nuclear losses. The majority of transmitted ions still experiences low nuclear losses, while some are subject to large nuclear losses. As can be seen in Fig. 2, fitting the convoluted simulation result to the full width of the experimental spectrum slightly overestimates the low-loss edge while underestimating the high-loss edge. The mismatch of the slopes of experiment and simulation particular at the highest losses means that nuclear losses do not fully account for the asymmetry of the random spectra. The corresponding trajectories, which are relatively rare and feature high energy losses, indicate a correlation between nuclear and electronic straggling, in particular as the asymmetry increases with decreasing ion energy. In other words, this observation implies that heavy ions at the energies investigated in our study predominantly experience electronic losses correlated with close-collision nuclear losses.

Other correlation effects such as bunching and packing effects on the electronic straggling [4,41] may also come into play. Whereas the bunching effect is relevant for inhomogeneous electron systems, and is caused by the spatial distribution of electrons in each atom, packing describes the correlation of electrons in different atoms [42]. Bunching is typically important for projectile energies close to the maximum of the stopping power and depends on the impact-parameter dependent mean energy loss. We have estimated the strength of bunching within the framework of the impact parameter dependent energy loss from the CasP program [43], but have found it to be negligible in the velocity regime at hand.

*Velocity scaling of straggling*

For all studied ions, energy-loss straggling along random trajectories is much larger than along channelled ones but significant differences in velocity scaling and comparison to theoretical predictions are observed. While straggling mostly increases with velocity for all channelled ions and H and He along random trajectories, for heavier ions and random trajectories, a different behaviour can be observed: the data uncorrected for nuclear losses becomes increasingly asymmetric with decreasing velocity leading to an increase of the straggling towards lower velocity (cf. Fig. 3). Comparison with Monte-Carlo calculations shows, however, that nuclear and geometrical straggling directly account only for a part of the observed increase. Even after subtraction of these contributions, the random straggling for B and Si ions does not decrease in the same way the channelled data does. Instead, the straggling seems to stay constant or even rise again with decreasing ion velocity. Such a trend can be explained by increasing contributions of relatively large energy transfers in individual collisions to the total energy loss. To further investigate this difference between light and heavier ions, a comparison with theoretical models is helpful.

Simple linear FEG model such as the Chu model cannot adequately describe straggling in Si at medium energies even for He. For He in random geometry, however, we find a good agreement between our full-peak data and the Penn-TCS model. A similar success of the Penn-TCS model was found in [36], where it could accurately predict the straggling of H in Pt. The case lies differently for B as an example of a heavier ion, for which none of the models could even predict the observed velocity scaling. The same is true for Si projectiles, where the simulated straggling in addition is much lower than the experimental data. Whereas the Penn-TCS model more adequately depicts the structures of the solid and the projectile, it still simulates electron-hole pair excitations. Note that in an FEG, straggling is much less sensitive to the projectile charge state than electronic stopping. As an example, the Penn-TCS calculations for $He^+$ and $He^{2+}$ are almost identical. The comparison between experimental straggling data for light ions and the Penn-TCS and Chu models shows that in this case, energy is indeed largely transferred via electron-hole pair excitations but that the electronic structure has to be taken into account. The fundamental disagreement between the model and the experimental data for heavier ions is a strong indication that individual large energy transfer happens via electron-promotion and charge-exchange events not electron-hole pair excitations. The analysis of electronic straggling allows for revisiting the results from [17,18], namely the trajectory dependence of electronic excitations and significantly higher electronic stopping in random geometry for all ions heavier than protons. For He this effect can be attributed mainly to more efficient electron-hole pair excitations caused by a higher mean charge state. For B and Si on the other hand, local energy deposition due to electron promotion, i.e., an electronic loss channel correlated with nuclear energy losses, contributes significantly to the total energy loss. Hereby, transitions can probably include one or several electrons as observed at lower ion energies [44,45].

**Summary and outlook**

We have analysed the electronic straggling of keV H, He, B and Si ions transmitted through Si nanomembranes both for random and channelling trajectories. For B and Si, we have accounted for nuclear straggling contributions with the help of Monte Carlo simulations. As for electronic stopping, we observe a strong trajectory dependence resulting in higher electronic straggling values for random geometry. Hereby, straggling seems to be even more sensitive to the trajectory, and differences between the two geometries are higher than for stopping.

By analysing electronic straggling in addition to energy loss, especially in combination with theoretical models, information on the specific type of involved energy-loss events can be gained. We find very direct evidence for increasing contribution of local losses for B and Si ions at low ion velocities visible as a plateau in the straggling-ion velocity curve. A theoretical model that accurately predicts all losses by electron-hole pair excitations, e.g., benchmarked against data at higher ion velocity, could consequently be compared to low-velocity data to estimate the amount of energy directly lost in charge-exchange events. By combining straggling data from experiment and theory in this way, charge exchange could be quantified.

## Acknowledgements

Support of accelerator operation by the Swedish Research Council VR-RFI (Contract No. 2017-00646_9 and 2019_00191) and the Swedish Foundation for Strategic Research (Contract No. RIF14-0053) is gratefully acknowledged.